\documentclass{article}

\usepackage[preprint]{neurips_2023}

\usepackage[utf8]{inputenc} 
\usepackage[T1]{fontenc}    
\usepackage[hidelinks]{hyperref}       
\usepackage{url}            
\usepackage{booktabs}       
\usepackage{amsfonts}       
\usepackage{nicefrac}       
\usepackage{microtype}      
\usepackage{xcolor}         
\usepackage[pdftex]{graphicx}
\usepackage{subcaption}
\usepackage{wrapfig}
\usepackage{float}
\setcitestyle{square,comma,numbers}

\title{AntiBARTy Diffusion for Property Guided Antibody Design}

\author{%
  Jordan Venderley \\
  Eli Lilly and Company\\
  Indianapolis, IN 46285\\
  \texttt{venderley\_jordan@lilly.com} \\
}

\begin{document}

\maketitle

\begin{abstract}

  Over the past decade, antibodies have steadily grown in therapeutic importance thanks to their high specificity and low risk of adverse effects compared to other drug modalities. While traditional antibody discovery is primarily wet lab driven, the rapid improvement of ML-based generative modeling has made in-silico approaches an increasingly viable route for discovery and engineering. To this end, we train an antibody-specific language model, AntiBARTy, based on BART (Bidirectional and Auto-Regressive Transformer) and use its latent space to train a property-conditional diffusion model for guided IgG de novo design. As a test case, we show that we can effectively generate novel antibodies with improved in-silico solubility while maintaining antibody validity and controlling sequence diversity.

\end{abstract}

\section{Introduction}

Industrial antibody discovery traditionally relies on phage display-based libraries or hybridoma technology using transgenic mice in order to obtain target-specific binders for antigens of interest.
Once isolated and sequenced, these high-affinity antibodies are used as a starting point for lead optimization in which properties important for e.g. safety, developability, manufacturability, etc. are also optimized while attempting to either maintain or further improve affinity.
While advancements in next-generation sequencing (NGS) technology have enabled high-throughput sequence determination, different approaches for sequencing/sorting come with tradeoffs between information completeness and throughput. \citep{Ho2018,Raybould2021}
In practice, despite throughput limitations, hybridoma-derived, single-cell sorted antibodies constitute most of the candidates selected from discovery for further engineering and much of the diversity offered by e.g. bulk repertoires is often under-sampled or not fully utilized.

Recent progress in machine learning is well-poised to take advantage of the wealth of data offered by NGS. 
By training on large sequence corpora, generative models can build strong priors on sequence space that can then be effectively sampled.
The ability to condition this sampling on biophysical properties, targets, or intra-complex chains offers a powerful route for augmenting the discovery process.
Since property data is often limited in volume, it's advantageous to construct this prior distribution via pretraining and then guide the sampling process towards favorable property modes by bootstrapping another generative model onto the frozen latent space.
This gives rise to the following strategy in which we: 
\begin{enumerate}
    \item Train an antibody-specific language model, AntiBARTy, based on BART (Bidirectional and Auto-Regressive Transformer) on large antibody sequence corpora
    \item Use its latent space to train a property-conditional diffusion model for classifier-free guided IgG de novo design
\end{enumerate}

Collectively, we refer to this approach as AntiBARTy Diffusion and demonstrate that it can effectively generate novel antibodies with improved (in-silico) properties.

\section{Related Work}

\subsection{Language Models}

With the availability of massive antibody sequence databases such as the Observed Antibody Space (OAS) and the wide-spread success of transformers in language modeling \cite{Vaswani2017}, it's unsurprising that numerous works have trained transformer-based models for antibodies. Typically these fall into two categories: encoder-only architectures for representation learning \cite{Ruffolo2021,Leem2022,Vig2021,Olsen2022,Vashchenko2022,Prihoda2022,Burbach2023} and decoder-only architectures for generative modeling \cite{Shuai2021}. 
BART (Bidirectional and Auto-Regressive Transformers) offers a less traditional approach and generalizes the encoder-only and decoder-only architectures. \cite{Lewis2020}  It conceptualizes pre-training as a decorruption task in which corrupted sequences are fed to the encoder and decorrupted with the decoder. These corruptions can be arbitrary, but canonically consist of token masking, deletion, insertion, and infilling and in the large corruption limit, BART reduces to a pure (decoder-only) language model. 

\subsection{Diffusion Models}


Diffusion models \cite{Ho2020} have risen to prominence through their SOTA performance in image generation and their impressive conditioning mechanisms.\cite{Rombach2022,Ramesh2022,Saharia2022} 
In protein design, structure-based SE(3)-equivariant diffusion models have seen great experimental success designing strong de novo binders without any experimental optimization e.g. mutagenesis. \cite{Watson2023} Joint sequence and structure-based diffusion has also been proposed for general proteins. \cite{Lisanza2023} While some diffusion models have been explored for antibodies \cite{Luo2022}, experimental success is much more difficult in this domain due to the sparsity of publicly available antibody structures and the low-resolution but high sequence variability of the complementarity-determining region (CDR) loops that drive antigen binding.

Unlike structures, there are billions of antibody sequences publicly available. \cite{Olsen2022-OAS} Melding the guidance capabilities of diffusion models with strong transformer-based priors on sequences offers the potential for computationally cheap and effective conditioning. Diffusion in the latent space of BART has been previously explored for natural language modeling, but to our knowledge not for any protein related tasks. \cite{Lovelace2022} Compared to this work, we propose the use of pooled embeddings for conditioning which we found yielded better downstream sequence quality and also leverage classifier-free guidance during sampling.

For sequence-based diffusion of antibodies, a recent work proposed a form of controllable, categorical diffusion with Bayesian optimization for fixed-length sequence design using an allotted edit budget. Excitingly, it reports experimental success in improving antibody binding affinities as part of lead optimization \cite{Gruver2023}. While our work, AntiBARTy Diffusion, was not designed to work within a fixed-edit budget, it does allow for variable sequence length guidance. Since indels play an important role in affinity maturation it is natural to include them in the design process, and they are a native component of AntiBARTy through indel-type corruption. \cite{DeWildt1999} Additionally our use of classifier-free guidance does not require an explicit discriminator which can in some cases be difficult to train. \cite{Ho2022}

\section{Methods}

\subsection{AntiBARTy}
We train a BART-style transformer \cite{Lewis2020} on all human IgG heavy and light sequences extracted from the Observed Antibody Space, totaling 254M heavy chains Fvs and 342M light chain Fvs. \cite{Olsen2022-OAS} In preprocessing we remove chains outside the range of [100,140] amino acids as well as any chains with unknown (X) amino acids. For added diversity, we augment this dataset with 28M similarly preprocessed sequences from UniProtKB .\cite{Bateman2023} All sequences are prepended with either a <heavy>, <light>, or <protein> tag and appended with an <EOS> token.

Both encoder and decoder contain 6 attention layers with the model totaling roughly 16M parameters.
Sequence corrupution (masking, indels, infilling) is performed on-the-fly during training, in alignment with best practices. \cite{Liu2019}
For accelerated performance we employ Flash Attention \cite{Dao2022} and use torch's DistributedDataParallel to parallelize across 4 Nvidia A100 GPUs. We train for 6 epochs over the course of 6 days.

\begin{wrapfigure}{r}{0.4\textwidth}
  \vspace{-5pt}
  \begin{center}
    \includegraphics[width=\linewidth]{"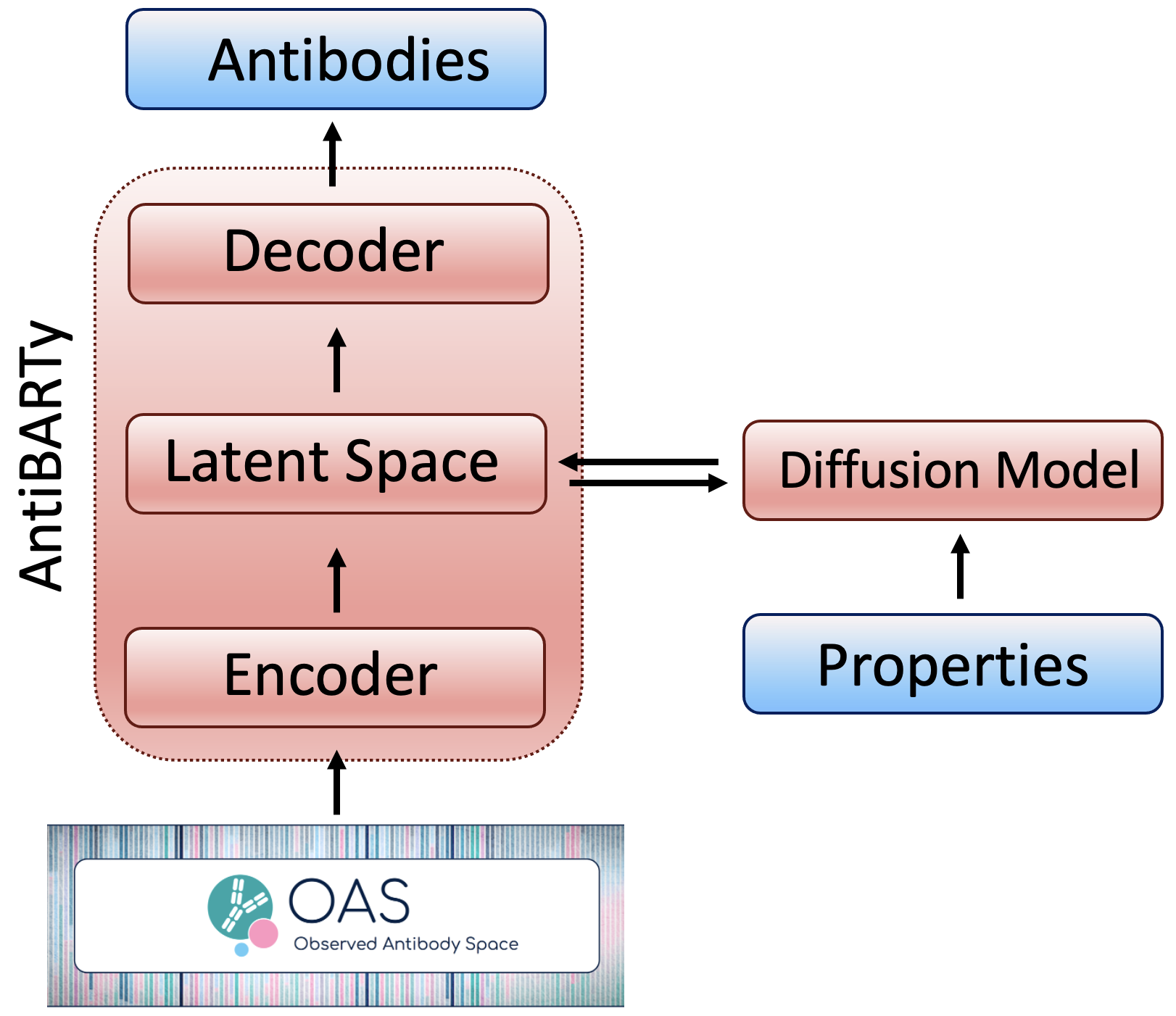"}
  \end{center}
  \caption{Schematic of the architecture used for AntiBARTy Diffusion}
  \label{fig:architecture}
  \vspace{-10pt}
\end{wrapfigure}

In order to mitigate known quality issues with OAS (e.g. missing portions near the N-terminus), after pretraining we briefly fine-tune our model on the paired subset of OAS which is of higher quality. This effectively realigns the distributional properties of the model and promotes valid antibody generation.
Importantly, during fine-tuning we also modify the cross-attention mechanism to operate on max-pooled embeddings over the sequence dimension of the encoder output. This deviates from previous work that addressed the variable length nature of natural language by sampling from empirical length distributions during inference.\cite{Lovelace2022} For our case, we found that the use of pooled embeddings led to improved sequence quality during sampling especially when used in conjunction with diffusion-generated encodings. After fine-tuning our language model, we freeze the parameters.

As an aside, we note that the autoregressive nature of BART gives us access to exact likelihoods instead of the quasi-likelihoods of encoder-only architectures like BERT. This is especially advantageous in the computation of evo-velocities for repertoire analysis since it can properly handle correlated indels and mutations. \cite{Ruffolo2021,Hie2022} We will explore this in future work.

\subsection{AntiBARTy Diffusion}

Once the language model has been frozen, we work with the normalized, continuous latent space offered by the encoder to train a property-conditional denoising diffusion probabilistic model (DDPM) \cite{Ho2020}. Our diffusion model uses a U-Net backbone  \cite{Ronneberger2015} with a depth of 3 layers and fuses the autoencoded input with learned class and time embeddings during upsampling. It contains roughly 3M parameters.

We jointly train the conditional and unconditional model using the AntiBARTy embeddings from the previous AntiBARTy fine-tuning set (unconditional) and those from a property dataset (conditional) of in-silico solubility scores calculated using Protein-Sol \cite{Hebditch2017} for a subset of VH chains in the paired OAS subset. We identify low ($< 0.45$) and high ($>0.7$) solubility classes and use roughly 20k samples from each class for training. During training we mask out the property class embeddings with a probability of 0.1 and to address the conditional/unconditional class imbalance we upsample the conditional training set by a factor of 10. A schematic of the full AntiBARTy architecture is provided in Fig. \ref{fig:architecture}. Once trained, we may effectively generate new antibodies by drawing a random latent vector from a standard multivariate normal distribution, denoising \`a la classifier-free guidance with a chosen guidance strength \cite{Ho2022}, and decoding with our AntiBARTy decoder. The decoder input is initialized with a <heavy> token. During decoding, we opted to greedily decode using Gumbel sampling with temperature 0.1. \cite{Jang2017} 

\section{Results}

\begin{figure}
     \centering
     \begin{subfigure}[b]{0.335\linewidth}
         \includegraphics[width=\textwidth]{"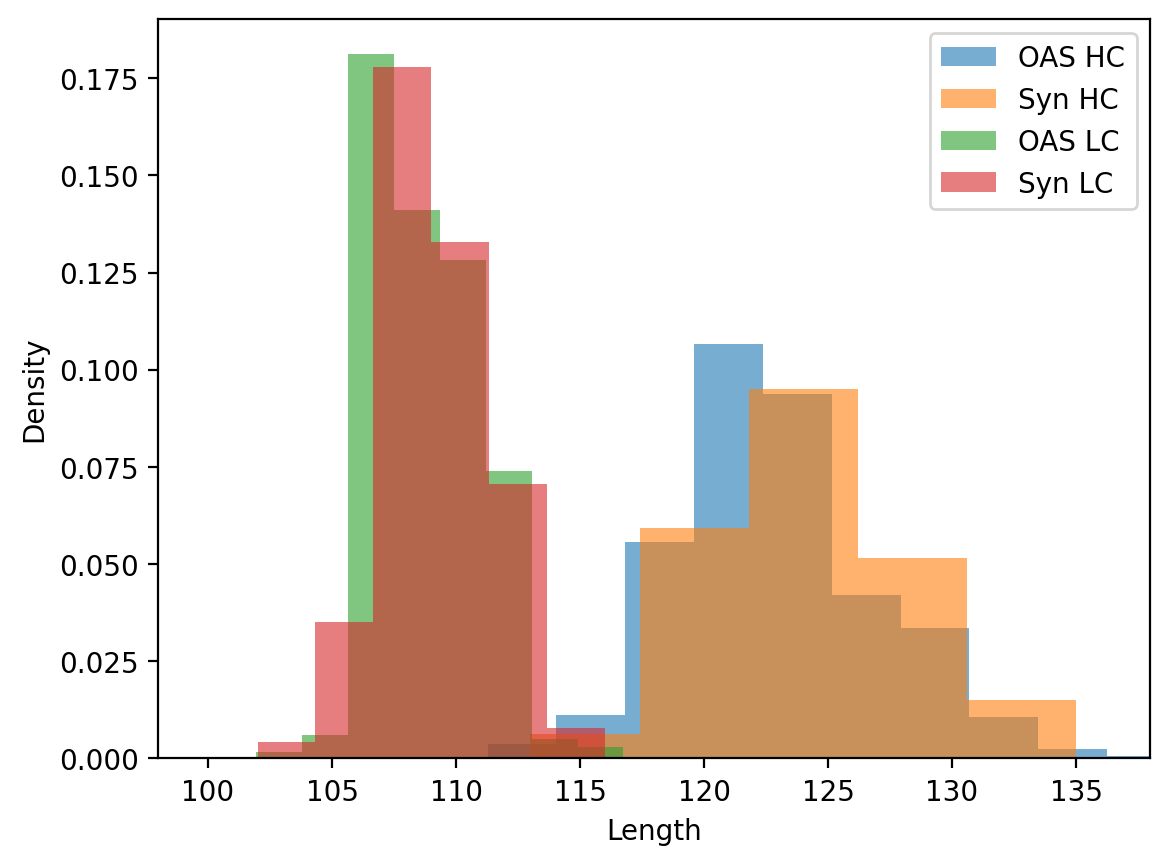"}
         \caption{}
         \label{fig:chain_conditional_len}
     \end{subfigure}
     \hfill
     \begin{subfigure}[b]{0.315\linewidth}
         \includegraphics[width=\textwidth]{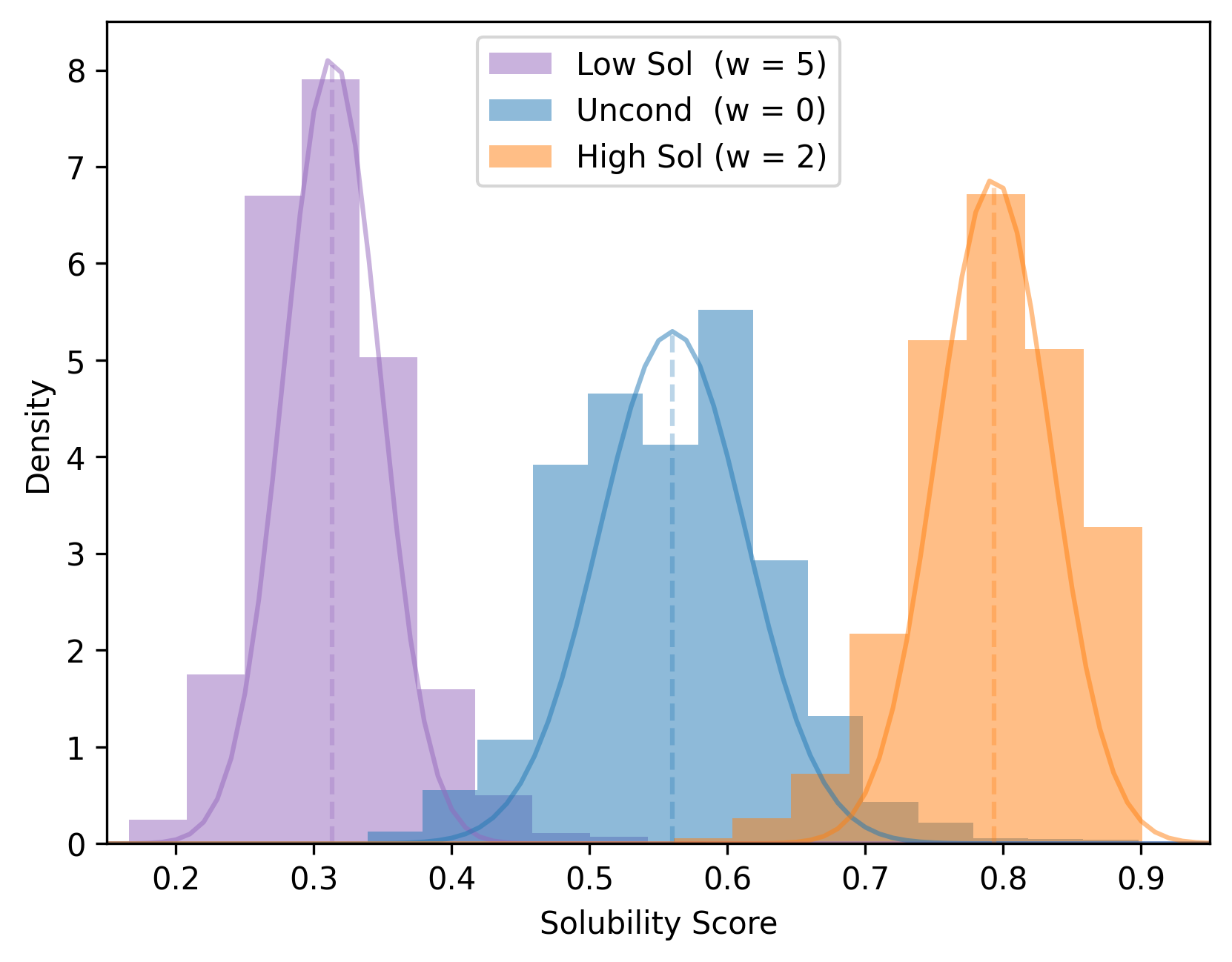}
         \caption{}
         \label{fig:guided_solubility}
     \end{subfigure}
     \begin{subfigure}[b]{0.33\linewidth}
         \includegraphics[width=\textwidth]{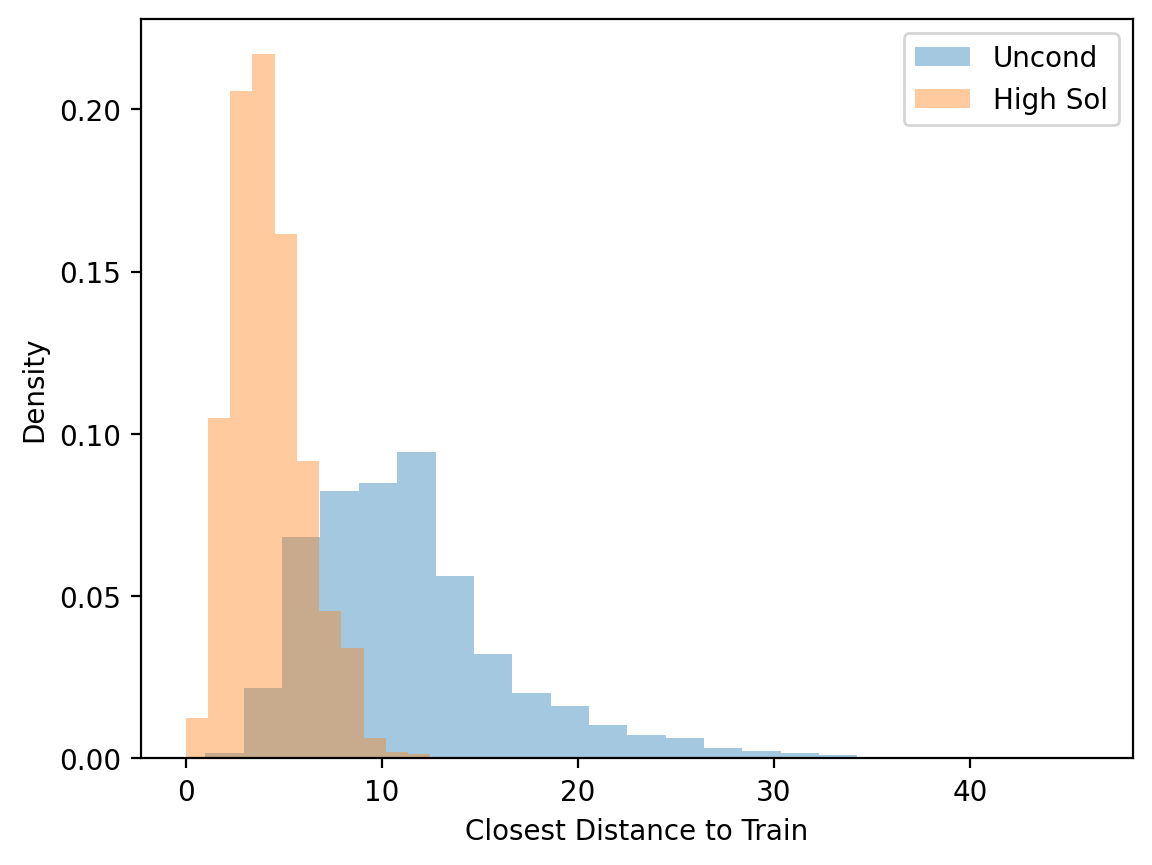}
         \caption{}
         \label{fig:closest_sequence}
     \end{subfigure}
     \caption{(a) Length distributions of heavy-chain conditional and light-chain conditional samples from AntiBARTy. (b) Distribution of unconditional and low/high-solubility conditional samples from AntiBARTy Diffusion. (c) Distribution of the distance of each AntiBARTy Diffusion sample to the closest sequence in the solubility training set. }
\end{figure}

To evaluate the quality of sequences generated by our language model and investigate the effect of fine tuning on the model, we fine-tuned a separate version of AntiBARTy to generate heavy or light chains only conditioning on the <heavy> or <light> token in the encoder. Sampling 1k sequences, we observe that the model can successfully recapitulate the statistics of the high quality training set despite the abundance of N-terminal truncations in the full dataset used during pretraining, see Fig \ref{fig:chain_conditional_len}. As a coarse quality check, we pass these sequences through ANARCI \cite{Dunbar2016} and observe that they are all classified appropriately as heavy or light chains with the distributions of assigned germline species matching that of the train set (predominately human but occasionally non-human for the light chains). All generated sequences are unique from each other and more detailed comparisons regarding distance from the train set are discussed in our evaluation of AntiBARTy Diffusion below.

\begin{figure}[H]
     \centering
     \begin{subfigure}[b]{0.48\linewidth}
         \includegraphics[width=\textwidth]{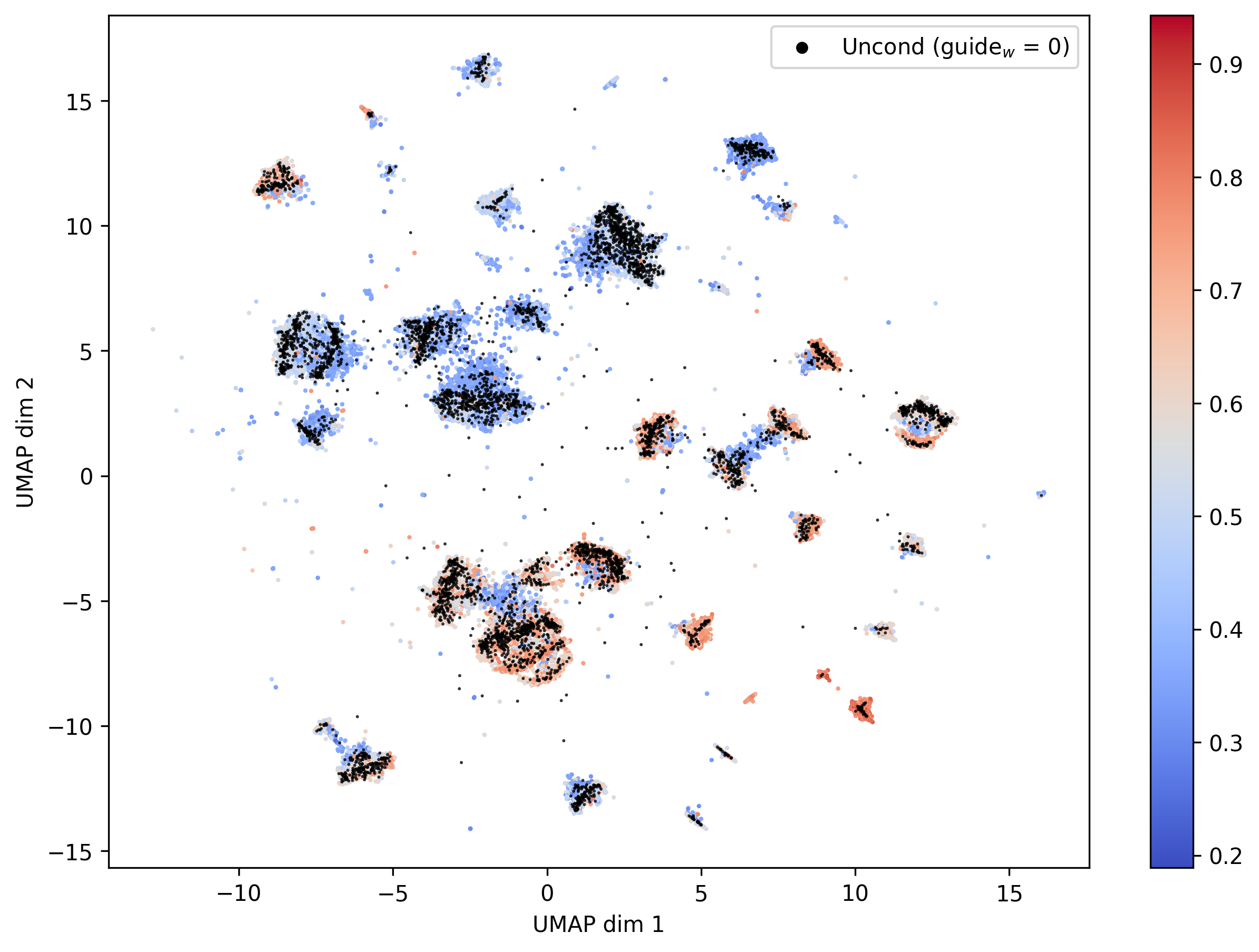}
         \caption{}
         \label{fig:uncond_umap}
     \end{subfigure}
     \begin{subfigure}[b]{0.48\linewidth}
         \includegraphics[width=\textwidth]{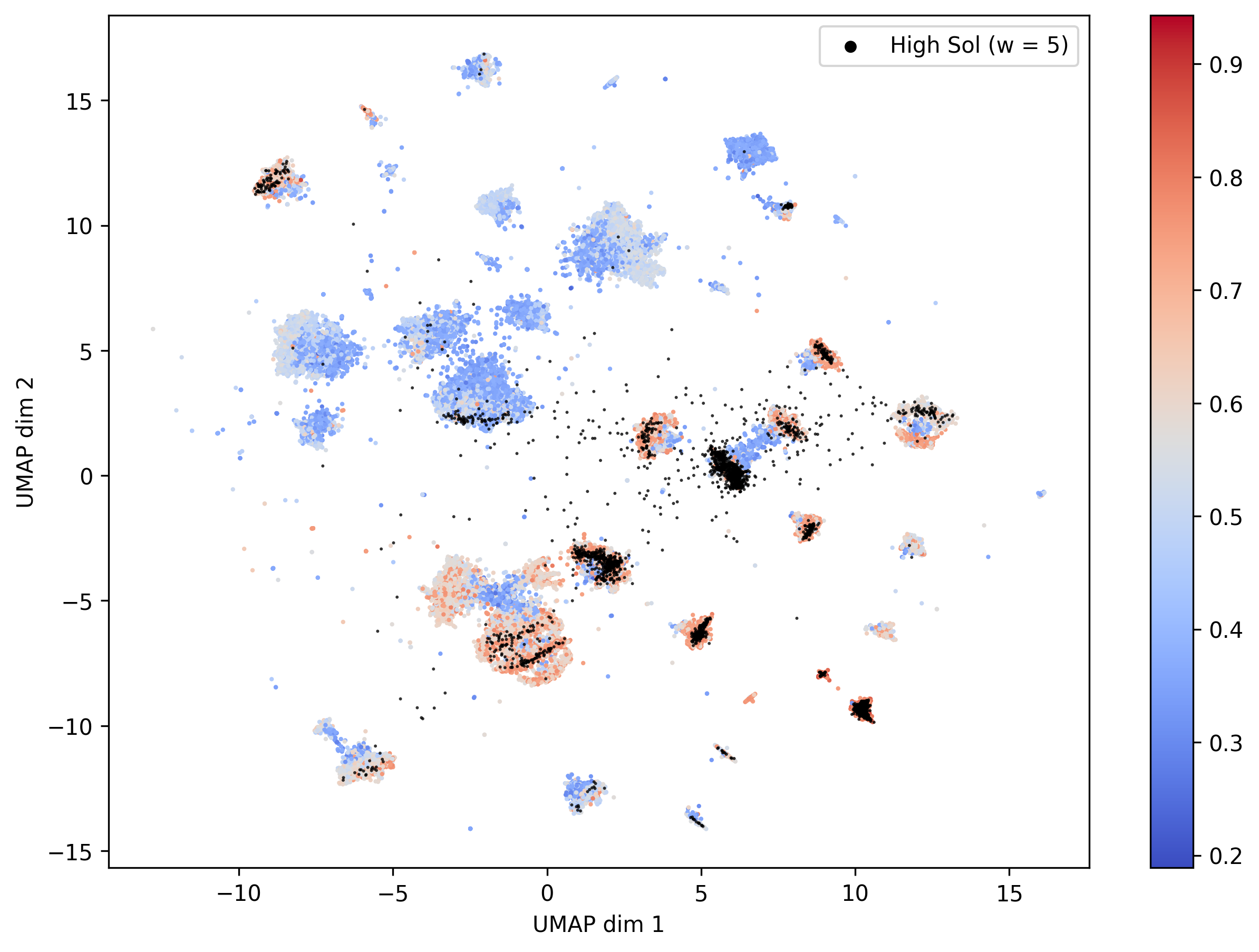}
         \caption{}
         \label{fig:high_umap}
     \end{subfigure}
     \caption{AntiBARTy Diffusion-generated embeddings of unconditional (a) and high solubility conditional (b) samples projected into 2D using UMAP. The training set is also shown and colored according to solubility.}
\end{figure}

To evaluate the property-guided de novo design capabilities of AntiBARTy Diffusion, we use it to generate low/high solubility sequences as determined by Protein-Sol \cite{Hebditch2017} using the approach described above. In Fig \ref{fig:guided_solubility} we plot the solubility distributions of 5k unguided and 5k guided low/high solubility class-conditional samples. The unconditional distribution matches that of the full train distribution (not shown) and low/high-solubility samples exhibit marked differences in solubility over the unconditioned distribution. Interestingly a larger guidance strength was required for the low-solubility generated samples in order to produce a shift in solubility of similar magnitude to their high solubility counterparts. ANARCI quality check confirms all synthetic sequences as heavy chains with more than 99.9\% assigned a human germline. Similarly more than 99.9\% of the generated samples are unique and don't already exist in the solubility training set. We subsequently focus on the high-solubility guided and unconditioned samples. For each sample, we find the closest sequence (in the Levenshtein-sense) from the solubility dataset and plot the distance distributions in Fig \ref{fig:closest_sequence}.  In order to explore mode coverage we use UMAP to embed our diffusion-generated encodings into 2D and plot the unconditional (Fig \ref{fig:uncond_umap}) and high-solubility guided (Fig \ref{fig:high_umap}) samples on top of the training set colored by solubility. We find that the high-solubility guided samples can successfully avoid the low-solubility modes and aggregate in the high solubility regions of the latent space.

\section{Conclusion}
We proposed AntiBARTy Diffusion for property-guided de novo antibody design and demonstrated its success in property guidance for in-silico solubilities. In future work, we plan to experimentally validate our approach and employ it for B-cell receptor repertoire analysis. We also plan to extend our method to enable infilling and the modification of existing antibodies using a combination of order-agnostic decoding \cite{Dauparas2022} and more sophisticated control mechanisms. \cite{Kawar2022,Zhang2023,Meng2022}




\newpage
\section{References}
\bibliographystyle{plainnat}
\bibliography{antibarty_refs}


\end{document}